# The SJR indicator: A new indicator of journals' scientific prestige

Borja González-Pereira[a], Vicente P. Guerrero-Bote[b] and Félix Moya-Anegón[c].

[a]SRG SCImago Research Group.
[b]University of Extremadura, Department of Information and Communication, Scimago Group, Spain.
[c]CSIC, CCHS, IPP, Scimago Group Spain.

**Abstract**

This paper proposes an indicator of journals' scientific prestige, the SJR indicator, for ranking scholarly journals based on citation weighting schemes and eigenvector centrality to be used in complex and heterogeneous citation networks such Scopus. Its computation methodology is described and the results after implementing the indicator over Scopus 2007 dataset are compared to an ad-hoc Journal Impact Factor both generally and inside specific scientific areas.

The results showed that SJR indicator and JIF distributions fitted well to a power law distribution and that both metrics were strongly correlated, although there were also major changes in rank. There was an observable general trend that might indicate that SJR indicator values decreased certain JIF values whose citedness was greater than would correspond to their scientific influence.

## *Introduction*

Citation analyses play an essential role in research evaluation systems, with their results being widely applied as complements to expert review.

The citedness of a scientific agent has for decades been regarded as an indicator of its scientific impact, and used to position it relative to other agents in the web of scholarly communications. In particular, various metrics based on citation counts have been developed to evaluate the impact of scholarly journals, one of which, the Impact Factor, has been extensively used for more than 40 years (Garfield, 2006).

However, recently there has emerged a new research trend aimed at developing impact metrics that consider not only the raw number of citations received by a scientific agent, but also the importance or influence of the actors who issue those citations (Palacios-Huerta, & Volij, 2004; Bollen, Rodríguez & van de Sompel, 2006; Ma et al., 2008; Bergstrom, 2007). These new metrics represent scientific impact as a function not of just the quantity of citations received but of a combination of the quantity and the quality.

The essential idea underlying the application of these arguments to the evaluation of scholarly journals is to assign weights to bibliographic citations based on the importance of the journals that issued them, so that citations issued by more important journals will be more valuable than those issued by less important ones. This "importance" will be computed recursively, i.e., the important journals will be those which in turn receive many citations from other important journals.

The first proposal in this sense in the field of Information Science was put forward by Pinsky & Narin (1976), with a metric they called "Journal Influence". Their proposed algorithm iterates the transfer of "prestige" from one journal to another until a steady-state solution is reached, whose values reflect the journals' scientific prestige. The "Journal Influence" indicator is a variant of the eigenvector centrality measure (Bonacich, 1987), with its calculation belonging to the group of eigenvector centrality methods in the domain of Network Theory. However, Pinsky & Narin's method presented problems in assigning centrality values to journals which were essentially related to the topological structure of the citation network.

With the arrival of the PageRank algorithm (Page et al., 1998) developed by the creators of Google, one had a computational model that resolved the aforementioned structure-related problems. Inspired in the Perron-Frobenius theorem, this algorithm modifies the network's structure by redefining the meaning of connections. In particular, it defines connections as the probability of going from one node to another, and, using a random-walker probabilistic model, transforms the citation network into a strongly connected graph, i.e., a network in which, given any two nodes, there is always some path to get from one to the other.

Applied to journal citation networks, this new model means that each connection between nodes (journals) represents the probability that a researcher, in documenting his or her research, goes from a journal to another selecting a random reference in a research article of the first journal. Values obtained after the whole process represent a "random research walk" that starts from a random journal to end in another after following an infinite process of selecting a random references in research articles. A random jump factor is added to represent the probability that the researcher chooses a journal by means other than following the references of research articles.

The method also defines an iterative algorithm that starts from certain initial pre-established values, and computes values of centrality until a steady-state solution is reached. The importance (prestige) of the nodes is redistributed at each iteration in terms of their connections with other nodes. The general formula used in this process is:

$$PR(Node_i, it_k) = \frac{1-\lambda}{N} + \lambda \sum_{j=1}^{N} (Connection_{(i,j)}) \cdot PR(Node_j, it_{k-1})$$

where the importance of node *i* in iteration *k* is set by the sum of the relative importance transferred by all the *i*-connected nodes. The amount of importance transferred by node *j* to node *i* is weighted by the strength of the connection between them, which is the fraction of references in node *j* in the year being considered that are to node *i*. The random jump factor, represented by the first term in the formula, is included to ensure convergence of the algorithm.

We here present an indicator of can be called "journal prestige" (Bollen, Rodríguez & van de Sompel, 2006), denominated the SCImago Journal Rank (SJR) indicator, that belongs to this new family of indicators based on eigenvector centrality. The SJR indicator is a size-independent metric aimed at measuring the current "average prestige per paper" of journals for use in research evaluation processes. It has already been studied as a tool for evaluating the journals in the Scopus database (Guz & Rushchitsky, 2009), compared with the Thomson Scientific Impact Factor (Falagas et al., 2008), and shown to constitute a good alternative for journal evaluation (Leydesdorff, 2009). In studying both bibliometric and usage indicators, Bollen, de Sompel, Hagberg, & Chute (2009) grouped the Impact Factor and the SCImago Journal Rank together, while clustering the Journal PageRank measure together with other "betweenness" centrality indicators. This was because the former are size-independent indicators rather than because they measure popularity as such.

In the following sections, we shall describe the methodological aspects of the development of the SJR indicator, and the results obtained with its implementation on the Elsevier's Scopus database, for which the data were obtained from the open access science evaluation resource SCImago Journal & Country Rank (2009).

## *Data*

We used Scopus as the data source for the development of the SJR indicator because it best represents the overall structure of world science at a global scale. Scopus is the world's largest scientific database. It covers all the journals included in the Thomson Reuters Web of Science (WoS) and more (Moya-Anegón et al., 2007; Leydesdorff, Moya-Anegón & Guerrero-Bote, 2010). Also, despite its only relatively recent launch in 2002, there are already various studies of its structure and coverage (Hane, 2004; Pickering, 2004; Jacso, 2004; LaGuardia, 2005). Indeed, its emergence has constituted competition for the WoS since it incorporates more services and data. The particular criteria that we analyzed in making our choice were selected as having a major influence on the final values of a bibliometric indicator, regardless of the methodological approach taken. They were the following:

1. Journal coverage.

2. Relationship between primary and secondary production in each journal of the database.

3. Assignment criteria for types of documents.

4. Accuracy of the linkage between references and source records.

Only documents published in 2007 included in the Scopus database were used for the main part of the study (in number, 1 821 744). All their references to documents present in the database in previous years were retrieved (in number, 22 370 409).

Documents are classified by area and category. There are 295 specific subject areas grouped into 26 subject areas. In addition, there is the General subject area containing multidisciplinary journals, such as Nature or Science. The subject areas are grouped into four categories on the Scopus "Basic Search" page (see the Scopus website, www.scopus.com, visited on 7 August 2009).

The four Scopus categories are:

- Life Sciences (> 4300 titles): Agricultural & Biological Sciences; Biochemistry, Genetics & Molecular Biology; Immunology & Microbiology; Neuroscience, Pharmacology, Toxicology & Pharmaceutics.

- Physical Sciences (> 7200 titles): Chemical Engineering; Chemistry; Computer Science; Earth & Planetary Science; Energy; Engineering; Environmental Science; Materials Science; Mathematics; Physics & Astronomy.

- Social Sciences (> 5300 titles):Arts & Humanities; Business, Management & Accounting; Decision Sciences; Economics, Econometrics and Finance; Psychology; Social Sciences.

- Health Sciences (> 6800 titles, including 100% coverage of Medline titles): Medicine; Nursing; Veterinary; Dentistry; Health Professions.

## *Method*

The SJR indicator is computed over a journal citation network where the nodes represent the scholarly journals in the database and the directed connections among the nodes the citation relationships among such journals. In our approach in particular, a directed connection between two journals is a normalized value of the number of references that the transferring journal makes to the recipient journal. The normalization factor used is the total number of references of the transferring journal in the year under study. The citation time window is set to three years, so that journal prestige is distributed among the references issued in the year under study directed to the papers published in the three previous years. The three-year citation window was chosen as the shortest one that embraces citation peaks of all the Subject Areas in Scopus as shown in Figure 1.

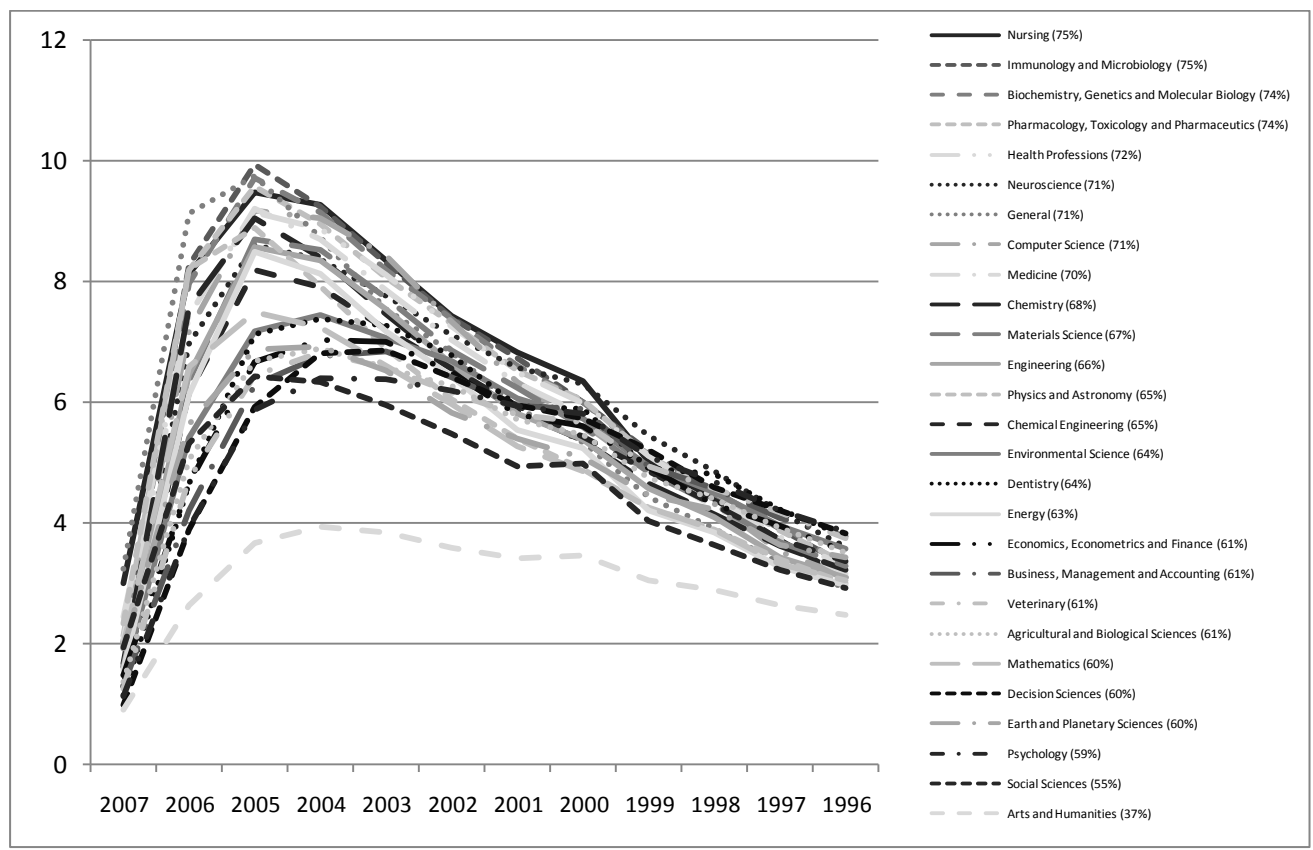

*Figure 1. Percentage for each of the last 12 years of the total references in articles published in 2007 and included in Scopus. In parentheses in the legend, the sum of these percentages.*

Next, in order to prevent excessive journal self-citation, the number of references that a journal may direct to itself is limited to a maximum 33% of its total references.

The computation is carried out using an iterative scheme that distributes prestige values among the journals until a steady-state solution is reached. The SJR algorithm begins by assigning an identical amount of prestige to each journal. Next, this prestige is redistributed in an iterative process whereby journals transfer their attained prestige to each other through the previously described connections. The process ends when the differences between journal prestige values in consecutive iterations do not surpass a pre-established threshold.

The SJR indicator is computed in two phases: the computation of the Prestige SJR (PSJR), a size-dependent measure that reflects the overall journal prestige; and the normalization of this measure to give a size-independent metric, the SJR indicator, which can be used to compare journals.

**Phase 1**

First, each journal is assigned the same initial prestige value 1/N, where N is the number of journals in the database. Then the iterative procedure begins. Each iteration assigns new prestige values to each journal in accordance with three criteria: (1) a minimum prestige value from simply being included in the database; (2) a publication prestige given by the number of papers included in the database; and (3) a citation prestige given by the number and "importance" of the citations received from other journals. The formula used for this calculation is the following:

$$PSJR_i = \overbrace{\frac{(1-d-e)}{N}}^{1} + \overbrace{e \cdot \frac{Art_i}{\sum_{j=1}^{N} Art_j}}^{2} + d \cdot \overbrace{\left[ \sum_{j=1}^{N} C_{ji} \cdot \frac{PSJR_j}{C_j} \cdot CF + \frac{Art_i}{\sum_{j=1}^{N} Art_j} \cdot \sum_{k \in DN} PSJR_k \right]}^{3}$$

**PSJRi** - Scimago Journal Rank of the Journal $i$.

**C$_{ji}$** - References from journal $j$ to journal $i$.

**C$_j$** - Number of references of journal $j$.

**d** – Constant: 0.9.

**e** – Constant: 0.0999.

**N** - Number of journals in the database.

**Art$_j$** - Number of primary items (articles, reviews, and conference papers) of journal $j$.

In the above formula, $e$ and $d$ are constants set to weight the amount of prestige that is achieved by means of publication and citation, respectively. Components 1 and 2, represented by the first two terms in the formula, are constant throughout the iteration, and together account for 10% of a journal's prestige value. Due to the complexity of Component 3, we will explain it in more detail.

The factor:

$$\sum_{j=1}^{N} C_{ji} \cdot \frac{PSJR_j}{C_j} \cdot CF$$

represents the prestige transferred to journal $i$ through the citations received from other journals. Each citation is weighted by the prestige achieved by the citing journal in the previous iteration divided by the number of references of any age found in that journal. Because only citations falling into the three-year window are used to distribute journal prestige, a procedure has to be defined to avoid losing the prestige value corresponding to the remaining citations in each iteration. To this end, a Correction Factor *CF* is introduced that spreads the undistributed prestige over all the journals proportionally to their accumulated prestige.

The formula for *CF* is:

$$CF = \frac{1 - \left(\sum_{k \in DN} PSJR_k\right)}{\sum_{h=1}^{N} \sum_{k=1}^{N} C_{kh} \cdot \frac{PSJR_k}{C_k}}$$

The denominator corresponds to the amount of prestige distributed through the citations falling in the three-year window, and the numerator is the amount of prestige available to be distributed, i.e., unity minus the prestige accumulated by the "dangling nodes" which will be explained in the next paragraph.

Finally,

$$\frac{Art_i}{\sum_{j=1}^{N} Art_j} \cdot \sum_{k \in DN} PSJR_k$$

distributes the prestige accumulated by the journals that do not cite other journals (called "dangling nodes", because they are not connected to any other node in the network) proportionally to the total number of primary items (articles, reviews, and conference papers) in the database.

The sum of the prestige values of all the journals in the database is normalized to unity in each iteration.

The iterative process terminates when the differences between the corresponding prestige values of all the journals in two consecutive iterations are no longer significant.

**Phase 2**

The Prestige SJR (PSJR) calculated in Phase 1 is a size-dependent metric that reflects the prestige of whole journals. It is not suitable for journal-to-journal comparisons since larger journals will tend to have greater prestige values. One needs to define a measure that is suitable for use in evaluation processes. To that end, the prestige gained by each journal, *PSJR*, is normalized by the number of primary items it has published (articles, reviews, and conference papers). Finally, these normalized *PSJR* values are increased proportionally to obtain an easy-to-use SJR indicator value. The procedure carried out in Phase 2 is given by the following formula:

$$SJR_i = c \cdot \frac{PSJR_i}{Art_i}$$

To put the methodological approach used to compute the SJR indicator in context, Table 1 presents a comparative synthesis of the principal differences between the SJR indicator, Eigenfactor.org's Article Influence, and Thomson Scientific's Impact Factor. We chose these two size-independent metrics for the comparison because of their extensive use as indicators in research evaluation.

*Table 1. Methodological differences between the SJR indicator, Article Influence, and Impact Factor.*

|  | SJR indicator | Article Influence | Impact Factor |
|---|---|---|---|
| General differences | | | |
| Source database | Scopus | Web of Science | Web of Science |
| Citation time frame | 3 years | 5 years | 2 years |
| Journal self-citation | Limited | Excluded | Included |
| Citation value | Weighted | Weighted | Unweighted |
| Specific differences | | | |
| Connections | Normalized by the total number of references in the citing journal | Normalized by the number of identified references in the citing journal | N.A. |

## *Statistical characterization*

We carried out a statistical characterization of the SJR indicator in order to contrast its capacity to depict what we have come to call the "average prestige per article" with journals' citedness per article. In the following paragraphs, we shall present comparisons of the rank distributions and scatterplots of the SJR indicator and the Journal Impact Factor, both overall for the entire database, and for some "subject areas" in different Scopus categories. We constructed an *ad hoc* JIF(3y) with a 3-year citation window so that any differences observed between the indicator values would be a consequence of the computation method and not of the time frame, citation window, etc. The study was performed for the year 2007 since its data can be considered stable. The data were downloaded from the SCImago Journal and Country Rank portal (http://www.scimagojr.com) on 8 November 2008.

Figure 2 shows a superposition of the overall SJR indicator and JIF(3y) value *vs* rank distributions. They are both similar to a power law distribution which would be represented in this semi-log plot by a descending, although steeper, straight line. The somewhat steeper fall-off of the SJR indicator distribution indicates that the prestige values are more concentrated, i.e., that there are fewer "prestigious" journals than highly cited ones. The two metrics are strongly correlated: their Spearman (rank) and Pearson correlation coefficients are 0.9248 and 0.8179, respectively. Generally, for the same journals, the SJR values are lower than the JIF(3y) values. Tables 2 and 3 give the statistical details of these statements.

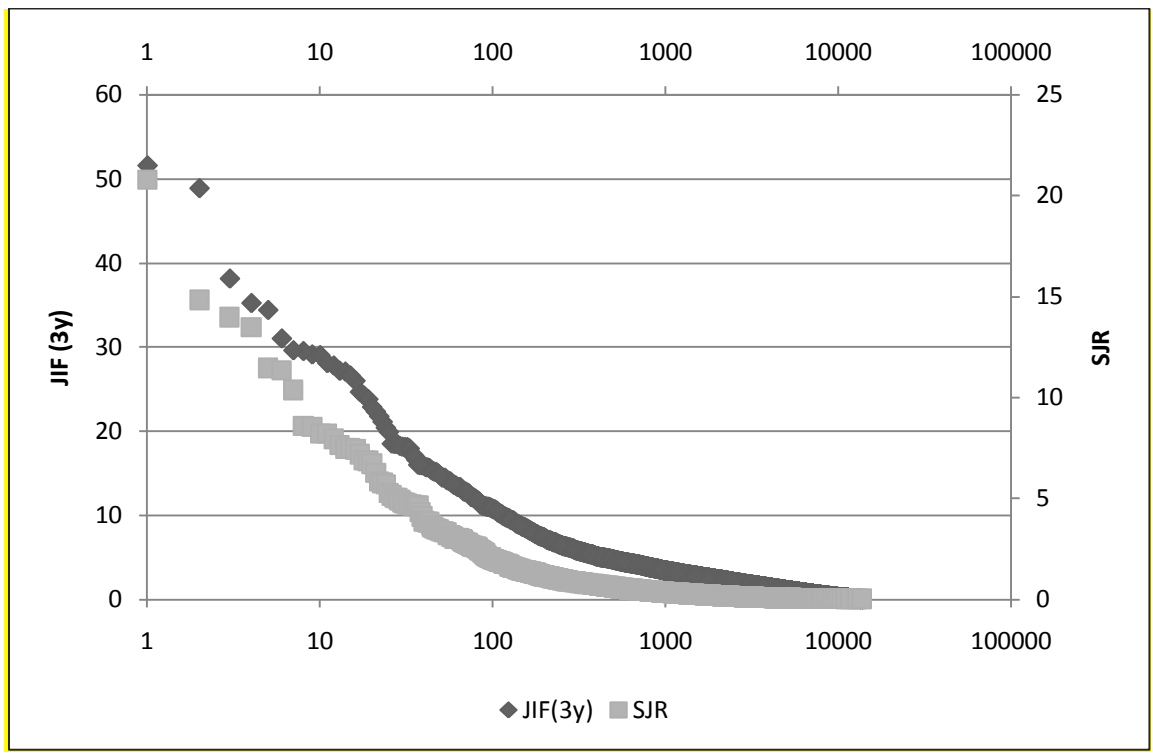

*Figure 2. Overlapping SJR indicator and JIF(3y) value-vs-rank distributions for the overall dataset.*

*Table 2. Average correlations of SJR indicator vs JIF(3y) by subject area and specific subject areas.*

|  | **Subject areas (27)** | **Specific subject areas (295)** |
|---|---|---|
| **Spearman** | x = 0.9316 ‖ sd = 0.0418 | x = 0.9117 ‖ sd = 0.1137 |
| **Pearson** | x = 0.8620 ‖ sd = 0.1108 | x = 0.8854 ‖ sd = 0.1413 |

*Table 3. Statistical parameters of the SJR indicator and JIF(3y) distributions.*

|  |  | Averages | Squared errors | Slope |
|---|---|---|---|---|
| Subject areas | SJR indicator | X = 0.1505 ‖ sd = 0.1716 | X = 1.2282 ‖ sd = 1.4219 | Sl = -1.3218 |
|  | JIF(3y) | X = 1.2681 ‖ sd = 0.7485 | X = 0.3918 ‖ sd = 0.6807 | Sl = -1.2178 |
| Specific subject areas | SJR indicator | X = 0.1376 ‖ sd = 0.1700 | X = 0.3070 ‖ sd = 0.7894 | Sl = -1.6561 |
|  | JIF(3y) | X = 1.2337 ‖ sd = 0.8237 | X = 0.1830 ‖ sd = 0.4853 | Sl = -1.3172 |

Figure 3 is a scatterplot of the same distributions as shown in Figure 2. One observes that the SJR indicator tends to lower the JIF(3y) rank of some journals, but not *vice versa*. Generally, this is the case with a journal that obtains many citations from relatively low importance journals, i.e., when the value of its centrality in the scientific discourse is lower than would be expected from its citedness.

The results presented in Table 4 serve to confirm the strong correlation between the two metrics. It lists the top ten journals in each metric and their corresponding ranks. Seven journals appear in both rankings, although their ranks differ.

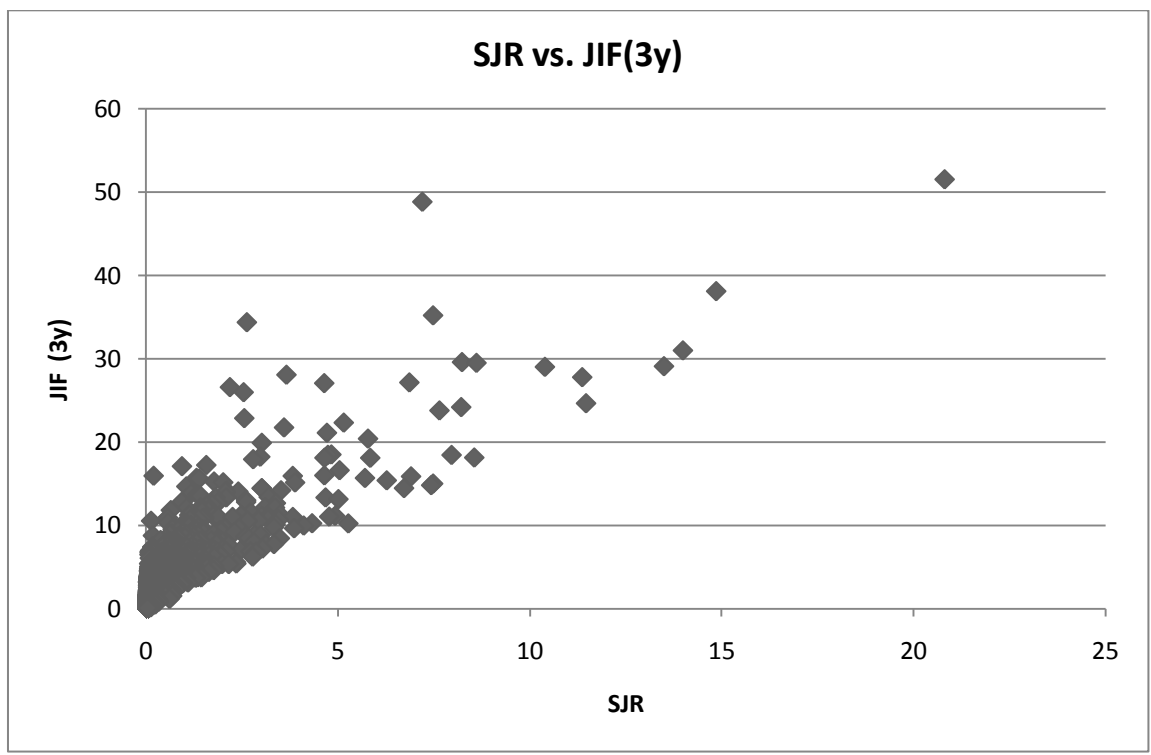

*Figure 3. Scatterplot of JIF(3y) vs the SJR indicator for the overall dataset.*

*Table 4: Top ten journals in the Scopus database, ranked by the SJR indicator and JIF(3y).*

| Title | SJR | Rank SJR | Rank (3y) | Title | IF(3y) | Rank SJR | Rank (3y) |
|---|---|---|---|---|---|---|---|
| Annual Review of Immunology | 20,81 | 1 | 1 | Annual Review of Immunology | 51,57 | 1 | 1 |
| Annual Review of Biochemistry | 14,86 | 2 | 3 | Ca-A Cancer Journal for Clinicians | 48,87 | 17 | 2 |
| Cell | 13,99 | 3 | 6 | Annual Review of Biochemistry | 38,12 | 2 | 3 |
| Annual Review of Cell and Developmental Biology | 13,5 | 4 | 9 | Physiological Reviews | 35,21 | 15 | 4 |
| Nature Immunology | 11,46 | 5 | 17 | Reviews of Modern Physics | 34,38 | 77 | 5 |
| Nature Reviews Molecular Cell Biology | 11,36 | 6 | 12 | Cell | 30,99 | 3 | 6 |
| Nature Reviews Immunology | 10,39 | 7 | 10 | Annual Review of Neuroscience | 29,59 | 10 | 7 |
| Nature Reviews Cancer | 8,61 | 8 | 8 | Nature Reviews Cancer | 29,49 | 8 | 8 |
| Immunity | 8,55 | 9 | 30 | Annual Review of Cell and Developmental Biology | 29,09 | 4 | 9 |
| Annual Review of Neuroscience | 8,23 | 10 | 7 | Nature Reviews Immunology | 29,02 | 7 | 10 |

In order to study the SJR indicator's behaviour in different scientific areas with distinct citation and publication patterns, we performed analyses involving several journal aggregations at the subject area and specific subject area levels. We shall describe three of these analyses corresponding to different Scopus categories.

First, we shall consider the Life Sciences specific subject area of Biochemistry, Genetics, & Molecular Biology (miscellaneous), which consists of 124 journals. Figure 4 shows the SJR indicator and JIF(3y) distributions, including the best fit regression straight lines, and Figure 5 shows the corresponding scatterplot. The Life Sciences category is characterized by a general concurrence of journal prestige and citedness, as is reflected by the strong correlations between the values of the SJR indicator and JIF(3y): 0.9486 for the Spearman rank correlation coefficient and 0.9533 for the Pearson correlation coefficient, both well above the overall category means given in Table 2. In sum, one can say that, in this area, highly cited journals receive a high ratio of citations from journals which are in turn highly cited.

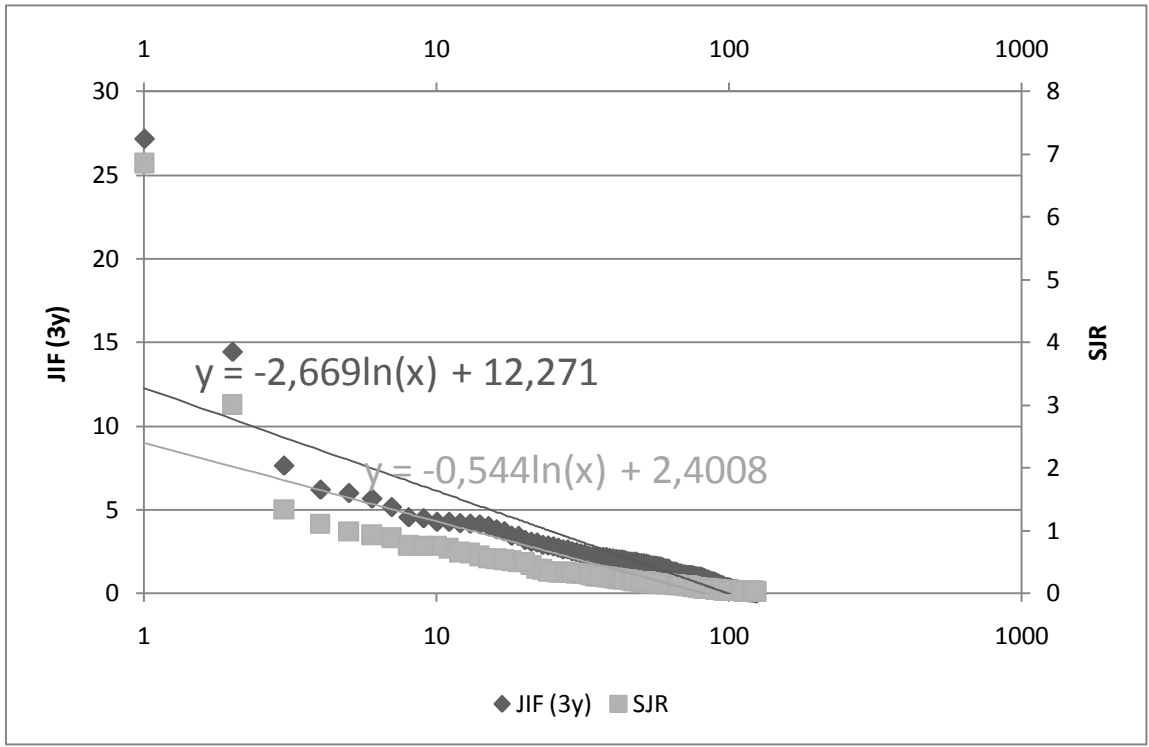

*Figure 4. Overlapping SJR indicator and JIF(3y) value-vs-rank distributions for the Biochemistry, Genetics & Molecular Biology (miscellaneous) Specific Subject Area.*

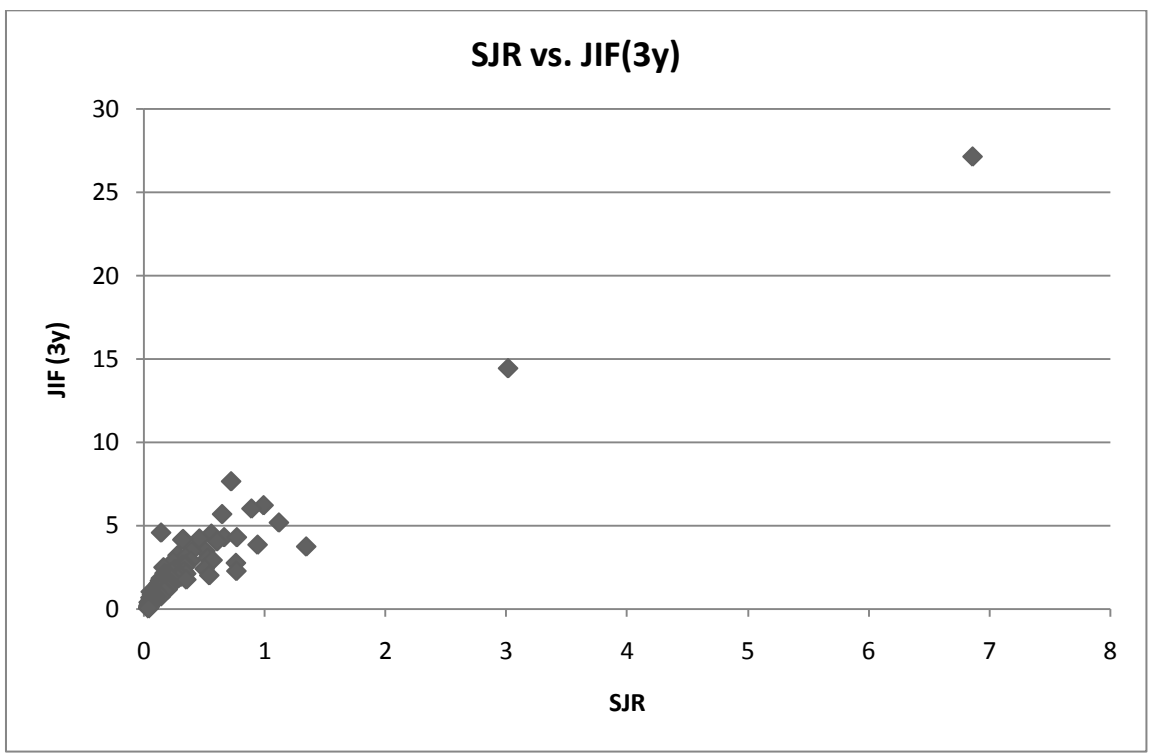

*Figure 5. Scatterplot of JIF(3y) vs the SJR indicator for the Biochemistry, Genetics & Molecular Biology (miscellaneous) Specific Subject Area.*

Table 5 shows five journals appearing in the top ten ranked by both the SJR indicator and JIF(3y). The differences between rankings were generally less than in the other comparative analyses of this study, except for the journal Trends in Glycoscience and Glycotechnology which was ranked 8th by JIF(3y) but 66th by the SJR indicator.

*Table 5: Top ten journals in the Scopus Specific Subject Area of Biochemistry, Genetics & Molecular Biology (miscellaneous), ranked by the SJR indicator and JIF(3y).*

| Title | SJR | Rank SJR | Rank (3y) | Title | IF(3y) | Rank SJR | Rank (3y) |
|---|---|---|---|---|---|---|---|
| Nature Medicine | 6,86 | 1 | 1 | Nature Medicine | 27,14 | 1 | 1 |
| Cytokine and Growth Factor Reviews | 3,01 | 2 | 2 | Cytokine and Growth Factor Reviews | 14,42 | 2 | 2 |
| Molecular systems biology [electronic resource]. | 1,34 | 3 | 17 | Natural Product Reports | 7,63 | 11 | 3 |
| Cellular and Molecular Life Sciences | 1,11 | 4 | 7 | Current Opinion in Lipidology | 6,19 | 5 | 4 |
| Current Opinion in Lipidology | 0,99 | 5 | 4 | Journal of Cellular and Molecular Medicine | 5,99 | 7 | 5 |
| Biochemistry and Cell Biology | 0,94 | 6 | 16 | Current Medicinal Chemistry | 5,66 | 13 | 6 |
| Journal of Cellular and Molecular Medicine | 0,89 | 7 | 5 | Cellular and Molecular Life Sciences | 5,15 | 4 | 7 |
| Molecular Cancer | 0,77 | 8 | 11 | Trends in Glycoscience and Glycotechnology | 4,55 | 66 | 8 |
| Acta Crystallographica Section D: Biological Crystallography | 0,76 | 9 | 33 | ChemBioChem | 4,49 | 16 | 9 |
| BMC Molecular Biology | 0,76 | 10 | 26 | Apoptosis | 4,28 | 12 | 10 |

Second, we shall consider a subject area in the Social Sciences category, in which it is known that the pattern of citation and publication is significantly different from that of the basic sciences (Nederhof, 2006). In particular, we analyzed the Psychology[1] subject area which comprises 334 journals. The results showed both indicators to closely follow a power law distribution (Figure 6), and to be strongly correlated with each other: Spearman rank correlation coefficient 0.9243, and Pearson correlation coefficient 0.8813, practically the same as the overall average values given in Table 2. In sum, one can say that there generally exists a strong correspondence between the notions of prestige and citedness in this subject area, although somewhat less so than in the previous specific subject area studied of Biochemistry, Genetics, & Molecular Biology (miscellaneous).

---

1 Psychology includes the Psychology (miscellaneous), Applied Psychology, Clinical Psychology, Developmental & Educational Psychology, Experimental & Cognitive Psychology, Neuropsychology & Physiological Psychology, and Social Psychology specific subject areas.

From the scatterplot in Figure 7, one observes that there were no marked variations in the two sets of values, although in some cases the SJR indicator tended to be lower than would have been expected from the JIF(3y) value.

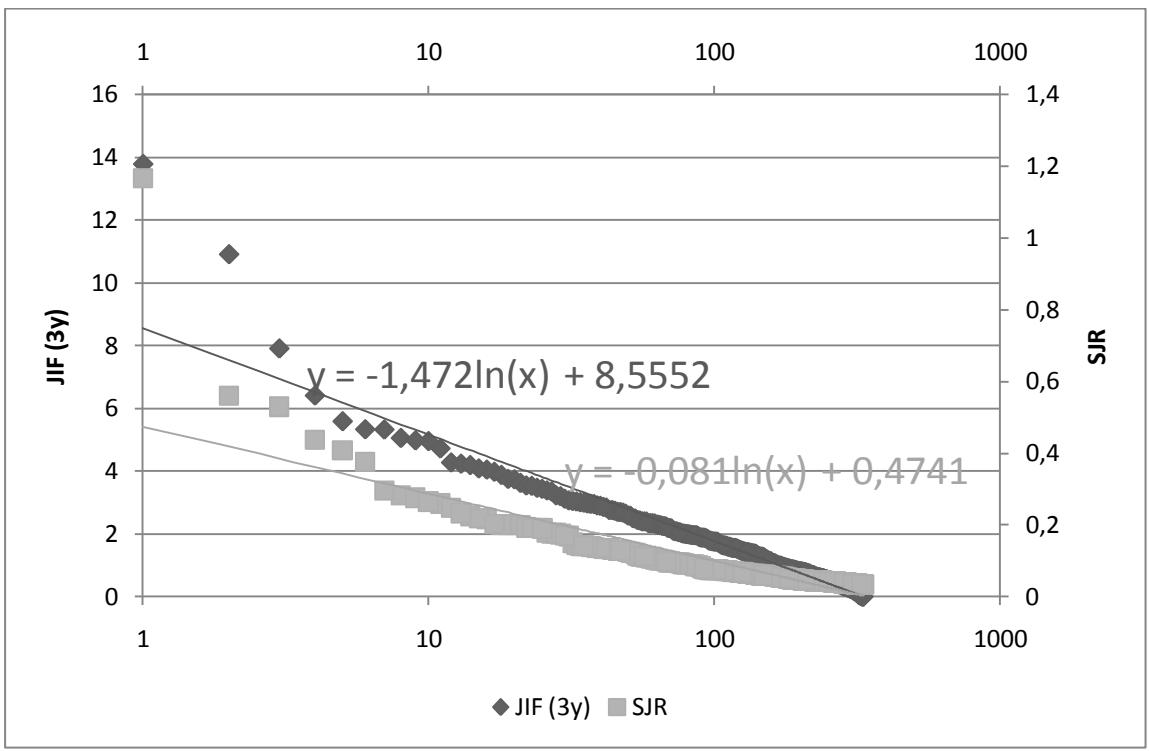

*Figure 6. Overlapping SJR indicator and JIF(3y) value-vs-rank distributions for the Psychology Subject Area.*

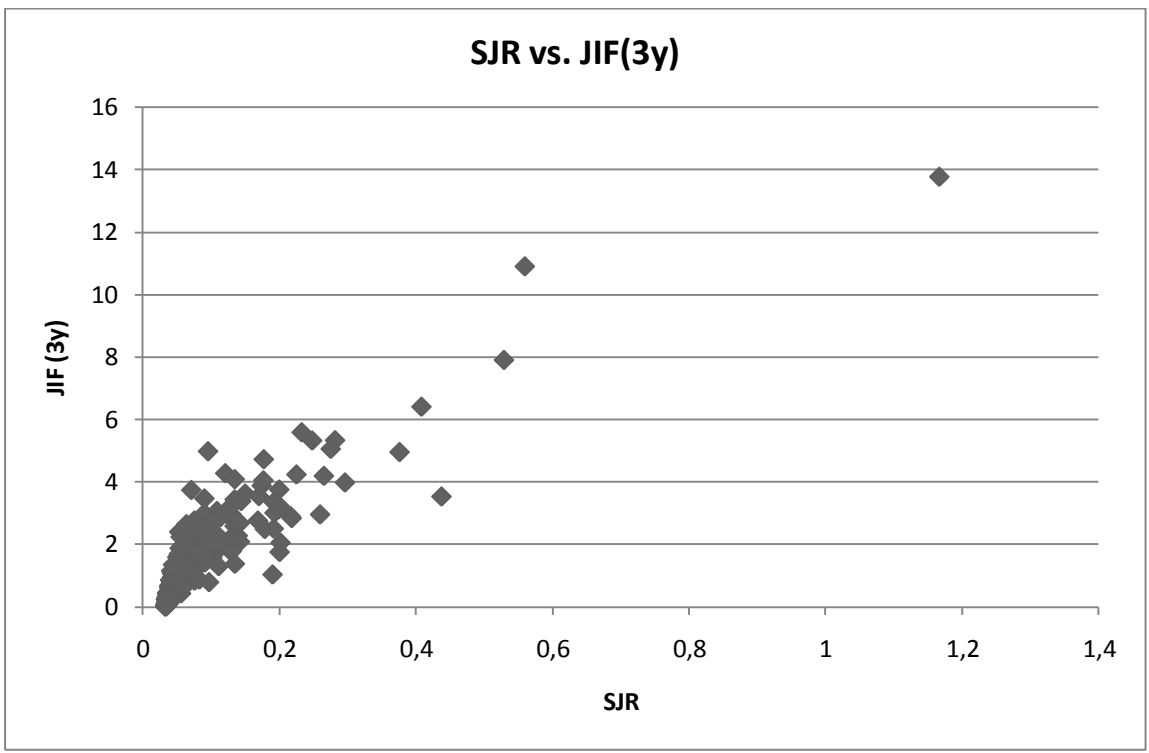

*Figure 7. Scatterplot of JIF(3y) vs the SJR indicator for the Psychology Subject Area.*

In Table 6, one observes that seven journals appeared in the top ten of both rankings, and that the first three positions coincided. There were no large rank changes except for the journal Educational Psychologist, which was ranked 9th by JIF(3y) but 73rd by the SJR indicator.

*Table 6: Top ten journals in the Scopus Subject Area of Psychology, ranked by the SJR indicator and JIF(3y).*

| Title | SJR | Rank SJR | Rank (3y) | Title | IF(3y) | Rank SJR | Rank (3y) |
|---|---|---|---|---|---|---|---|
| Annual Review of Psychology | 1,17 | 1 | 1 | Annual Review of Psychology | 13,77 | 1 | 1 |
| Psychological Bulletin | 0,56 | 2 | 2 | Psychological Bulletin | 10,9 | 2 | 2 |
| Psychological Review | 0,53 | 3 | 3 | Psychological Review | 7,9 | 3 | 3 |
| Personality and Social Psychology Bulletin | 0,44 | 4 | 23 | Journal of Experimental Psychology: General | 6,4 | 5 | 4 |
| Journal of Experimental Psychology: General | 0,41 | 5 | 4 | Journal of Abnormal Psychology | 5,58 | 13 | 5 |
| American Psychologist | 0,38 | 6 | 10 | Psychotherapy and Psychosomatics | 5,33 | 8 | 6 |
| Review of General Psychology | 0,3 | 7 | 17 | Psychological Science | 5,32 | 12 | 7 |
| Psychotherapy and Psychosomatics | 0,28 | 8 | 6 | Child Development | 5,05 | 9 | 8 |
| Child Development | 0,28 | 9 | 8 | Educational Psychologist | 4,98 | 73 | 9 |
| Personnel Psychology | 0,27 | 10 | 14 | American Psychologist | 4,95 | 6 | 10 |

Finally third, we shall describe our analysis of the Computer Science Subject Area[2] in the Scopus Physical Sciences category, comprising 578 journals. Technical domains such as Computer Science are known to exhibit singular publication and citation patterns which differentiate them from other areas of science (Moed, 2005). One observes in Figure 8 that here again both distributions closely followed a power law distribution. The correlations between the values of the two metrics were, however, were the lowest of those analyzed – 0.8644 for the Spearman rank correlation coefficient and 0.6479 for the Pearson correlation coefficient. The scatterplot (Figure 9) confirms this with the clearly large dispersion of the values, and one observes in Table 7 that there was only one journal in the top ten of both rankings. Indeed, there were striking changes in rank: for example, Annals of Mathematics and Artificial Intelligence was ranked 1st by the SJR indicator but 144th by JIF(3y). Clearly, in this subject area there is little connection between the notions of prestige and citedness.

---

2  This includes the specific subject areas: Computer Science (miscellaneous); Artificial Intelligence; Computational Theory & Mathematics; Computer Graphics & Computer-Aided Design; Computer Networks & Communications; Computer Science Applications; Computer Vision & Pattern Recognition; Hardware & Architecture; Human-Computer Interaction; Information Systems; Signal Processing; and Software.

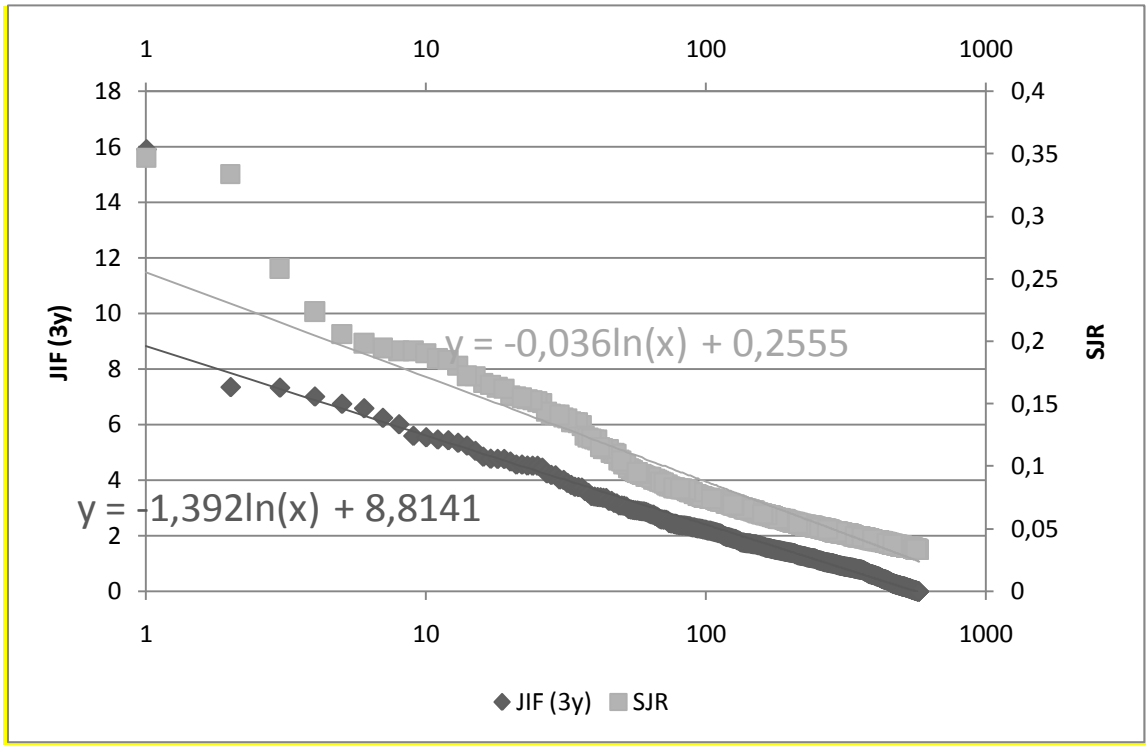

*Figure 8. Overlapping SJR indicator and JIF(3y) value-vs-rank distributions for the Computer Science Subject Area.*

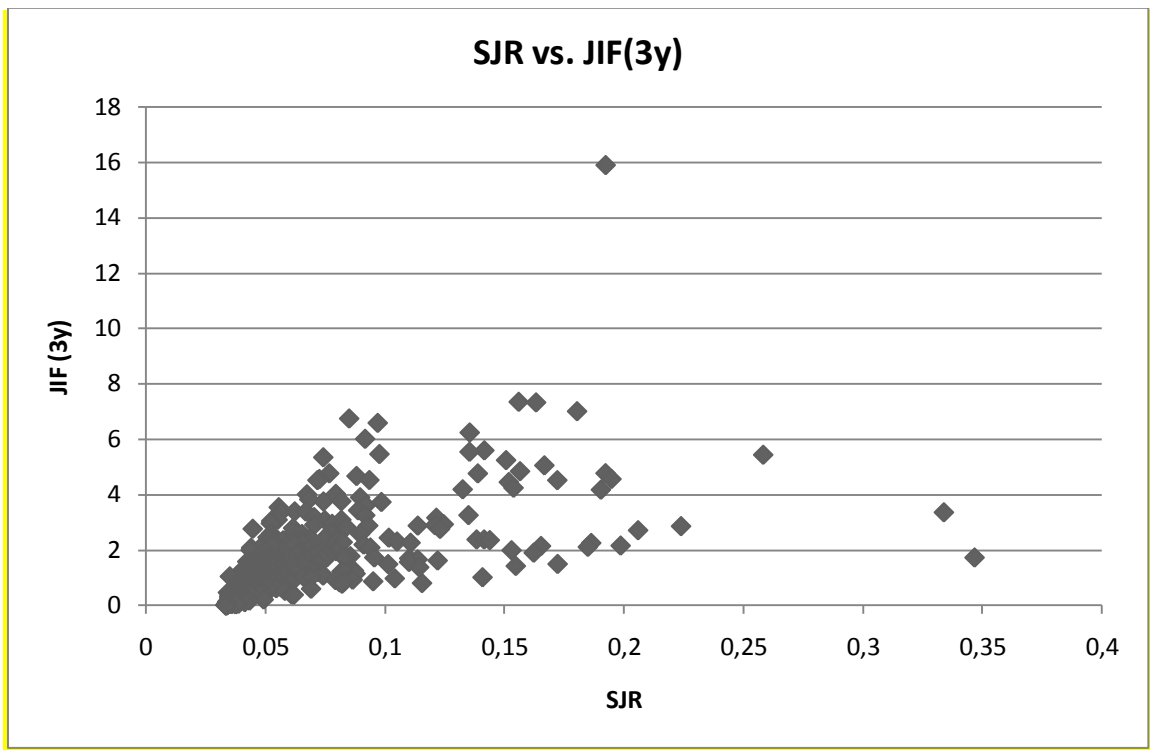

*Figure 9. Scatterplot of JIF(3y) vs the SJR indicator for the Computer Science Subject Area.*

*Table 7: Top ten journals in the Scopus Subject Area of Computer Science, ranked by the SJR indicator and JIF(3y).*

| Title | SJR | Rank SJR | Rank (3y) | Title | IF(3y) | Rank SJR | Rank (3y) |
|---|---|---|---|---|---|---|---|
| Annals of Mathematics and Artificial Intelligence | 0,35 | 1 | 144 | ACM Transactions on Information and System Security | 15,92 | 8 | 1 |
| Parallel Processing Letters | 0,33 | 2 | 44 | IEEE Transactions on Mobile Computing | 7,36 | 21 | 2 |
| Conference on Human Factors in Computing Systems - Proceedings | 0,26 | 3 | 12 | Proceedings of the Annual International Conference on Mobile Computing and Networking, MOBICOM | 7,34 | 18 | 3 |
| International Journal of Parallel, Emergent and Distributed Systems | 0,22 | 4 | 61 | Proceedings of the ACM Conference on Computer and Communications Security | 7,02 | 13 | 4 |
| Journal of Cases on Information Technology | 0,21 | 5 | 66 | ACM Transactions on Internet Technology | 6,76 | 73 | 5 |
| International Journal of Human Computer Studies | 0,2 | 6 | 106 | Medical Image Analysis | 6,6 | 55 | 6 |
| ACM/SIGDA International Symposium on Field Programmable Gate Arrays - FPGA | 0,19 | 7 | 21 | ACM Transactions on Computer-Human Interaction | 6,25 | 33 | 7 |
| ACM Transactions on Information and System Security | 0,19 | 8 | 1 | Proceedings of the ACM SIGPLAN Conference on Programming Language Design and Implementation (PLDI) | 6,02 | 63 | 8 |
| Proceedings of the Symposium on Interactive 3D Graphics | 0,19 | 9 | 17 | Proceedings of the ACM SIGSOFT Symposium on the Foundations of Software Engineering | 5,6 | 28 | 9 |
| Journal of Computing and Information Science in Engineering | 0,19 | 10 | 29 | Proceedings of the Conference on Object-Oriented Programming Systems, Languages, and Applications, OOPSLA | 5,55 | 34 | 10 |

## *Conclusions*

This study has presented the development of the SJR indicator, a new metric of the scientific influence of scholarly journals aimed at use in conventional processes of research evaluation.

Since it is constructed on the Scopus database, we believe it will best reflect the citation relationships among scientific sources. However, at the same time, it will be necessary to adapt the PageRank method of computation to the particularly complex and heterogeneously structured characteristics of such citation network.

Methodologically, the SRJ indicator establishes different values for citations according to the scientific influence of the journals that generate them. It uses a three-year citation window – long enough to cover the citation peak of a significant number of journals, and short enough to be able to reflect the dynamics of the scholarly communication process. It restricts a journal's self-citation to a maximum of 33% of its issued references, so that excessive self-citation will not involve artificially inflating a journal's value, but without touching the normal process of self-citation.

Technically, the method proposes a solution to the known computational issues of PageRank-based methods with respect to the existence of journals which have no references to other journals in the database. For this purpose, the solution we use is to distribute these journals' accumulated prestige values among all the other journals in the database proportionally to their number of published papers. We also propose that the normalization of the connections between the journals is by means of the total number of references found in the citing journal instead of considering only those falling within the citation window. This obviates the issue of what to do with journals that transfer their accumulated prestige through so few references that they have little statistical significance.

The statistical characterization of the SJR indicator and its comparison with an *ad hoc* constructed method, JIF(3y), which was based on the unweighted counts of citations, provided quite conclusive results. While there existed a strong overall correlation between a journal's citedness and its scientific influence in terms of eigenvector centrality, there were also major changes in rank. Although both approximations closely follow a power law distribution, scientific prestige is more concentrated in fewer journals.

There was an observable general trend that the SJR indicator values decreased certain JIF(3y) values. Subsequent studies of this trend would help one to determine whether, as we intuit to be the case, this pattern is due to the SJR indicator reducing the rank of journals whose citedness is greater than would correspond to their scientific influence.

In sum, the SJR is a bibliometric indicator that measures the prestige or influence of a scientific journal article, calculated with the largest and most nearly complete bibliographic database, and using a citation window of 3 years that is wide enough to include most of the citations, and dynamic enough to measure the evolution of scientific journals.

## *Acknowledgements*

This work was financed by the Junta de Extremadura - Consejería de Educación Ciencia & Tecnología and the Fondo Social Europeo as part of research project PRI06A200, and by the Plan Nacional de Investigación Científica, Desarrollo e Innovación Tecnológica 2008-2011 and the Fondo Europeo de Desarrollo Regional (FEDER) as part of research projects TIN2008-06514-C02-01 and TIN2008-06514-C02-02.

SCImago Journal and Country Rank. SCImago Research Group. Available at: http://www.scimagojr.com [Accessed: 9 October 2009].